\documentclass[11pt,a4paper]{article}
\usepackage{ae,aecompl}
\usepackage[LGR,T1]{fontenc}
\usepackage[latin9]{inputenc}
\usepackage{mathrsfs}
\usepackage{amsmath}
\usepackage{amssymb}
\usepackage{esint}

\makeatletter

\pdfoutput=1 

\usepackage{jheppub}

\usepackage{etoolbox}
    
\patchcmd{\maketitle}{\@fpheader}{}{}{}

\usepackage{aecompl}
\usepackage{amsfonts}
\usepackage{bbm}

\title{\boldmath Nonlinear automorphism of the conformal algebra in 2D and continuous $\sqrt{T\bar{T}}$ deformations }

\author[a]{David Tempo,}
\author[b,c]{Ricardo Troncoso}

\affiliation[a]{Departamento  de  Ciencias  Matem\'{a}ticas  y  F\'{i}sicas, Universidad  Cat\'{o}lica  de  Temuco,  Montt  56,  Casilla  15-D,  Temuco,  Chile.}
\affiliation[b]{Centro de Estudios Cient\'{i}ficos (CECs), Av. Arturo Prat 514, Valdivia, Chile.}
\affiliation[c]{Facultad de Ingenier\'{i}a, Arquitectura y Dise\~{n}o, Universidad San Sebasti\'{a}n, sede Valdivia,
General Lagos 1163, Valdivia 5110693, Chile}

\emailAdd{jtempo@uct.cl}
\emailAdd{troncoso@cecs.cl}
\emailAdd{ricardo.troncoso@uss.cl}

\preprint{CECS-PHY-22/05}

\abstract{The conformal algebra in 2D (Diff($S^{1}$)$\oplus$Diff($S^{1}$)) is shown to be preserved under a nonlinear map that mixes both chiral (holomorphic) generators $T$ and $\bar{T}$.
 It depends on a single real parameter and it can be regarded as a ``nonlinear $SO(1,1)$ automorphism.'' The map preserves the form of the momentum density and naturally induces a flow on the energy density by a marginal $\sqrt{T\bar{T}}$ deformation.
 In turn, the general solution of the corresponding flow equation of the deformed action can be analytically solved in closed form, recovering the nonlinear automorphism.
The deformed CFT$_{2}$ can also be described through the original theory on a field-dependent curved metric whose lapse and shift functions are given by the variation of the deformed Hamiltonian with respect to the energy and momentum densities, respectively.
 The conformal symmetries of the deformed theories can then also be seen to arise from diffeomorphisms that fulfill suitably deformed conformal Killing equations. Besides, Cardy formula is shown to map to itseft under the nonlinear automorphism. As a simple example, the deformation of $N$ free bosons is briefly addressed, making contact with recent related results and the dimensional reduction of the ModMax theory. Furthermore, the nonlinear map between the conformal algebra in 2D and its ultra/non-relativistic
versions (BMS$_{3}$$\approx$CCA$_{2}$$\approx$GCA$_{2}$), including the corresponding finite $\sqrt{T\bar{T}}$ deformation, are recovered from a limiting case of the nonlinear automorphism. The extension to a three-parameter nonlinear $ISO(1,1)$ automorphism
of the conformal algebra, and a discrete nonlinear automorphism of BMS$_{3}$ are also briefly discussed.}

\makeatother

\begin{document}
\maketitle \flushbottom \newpage{}

\section{Introduction }

Symmetries spanned by infinite-dimensional Lie algebras tightly constrain
the possibilities to devise theories or computing amplitudes, among
others. Along these lines, consistent nonlinear maps between the generators
of infinite-dimensional linear algebras can also be very useful, but
since they are just sporadically found, the systematic study of this
kind of mappings turns out to be uncharted territory either in physics
or mathematics (for a recent related discussion see \cite{Fuentealba:2022yqt}).
A time-honored example of this sort is the Sugawara construction \cite{Sugawara:1967rw},
which is known to play a very relevant role in the context of conformal
field theories (CFT's) in two-dimensional spacetimes. In this class
of mapping, the generators of the Virasoro algebra are obtained from
a precise quadratic combination of those of an affine Kac-Moody algebra
(see e.g. \cite{DiFrancesco:1997nk,Blumenhagen:2009zz}). Similar
quadratic Sugawara-like relationships, realized in a variety of setups,
are also known to exist for the ultra/non-relativistic limit of the
conformal algebra in 2D (CCA$_{2}$$\approx$GCA$_{2}$$\approx$BMS$_{3}$))
\cite{Barnich:2013yka,Donnay:2015abr,Afshar:2016wfy,Afshar:2016kjj,Grumiller:2019fmp},
and its supersymmetric extensions \cite{Barnich:2014cwa,Barnich:2015sca,Banerjee:2016nio,Fuentealba:2017fck,Banerjee:2019lrv,Banerjee:2021uxl}
whose generators also emerge from different quadratic combinations
of current algebras. 

Another type of nonlinear mappings of the class under discussion is
naturally formulated in the context of integrable systems. Indeed,
since the Virasoro algebra describes one of the Poisson structures
of the KdV hierarchy (see e.g., \cite{Das:1989fn,olver1993applications,Dunajski:2010zz}),
their generators are also nonlinearly related to the infinite-dimensional
Abelian algebra of conserved charges, spanned by the subset of commuting
generators of the enveloping algebra. A similar nonlinear relationship
of this kind is also known to hold for the generators of the BMS$_{3}$
algebra \cite{Fuentealba:2017omf}.

A different kind of nonlinear map that requires going beyond the enveloping
algebra has been recently introduced in \cite{Rodriguez:2021tcz},
which relates the generators of the classical (nonanomalous) conformal
algebra in 2D (Diff($S^{1}$)$\oplus$Diff($S^{1}$)) with those of
the BMS$_{3}$ algebra. Under this mapping, the Hamiltonian of a generic
CFT$_{2}$, with chiral (holomorphic) conformal generators given by
$T$ and $\bar{T}$, acquires a finite marginal non-analytic deformation
determined by $\sqrt{T\bar{T}}$, so that the deformed theory it is
no longer a CFT$_{2}$, but a conformal Carrollian field theory instead,
because its energy and momentum densities fulfill the BMS$_{3}$ algebra\footnote{These marginal deformations are clearly different from the well-known
irrelevant $T\bar{T}$ ones \cite{Zamolodchikov:2004ce,Smirnov:2016lqw,Cavaglia:2016oda},
whose diverse properties have been studied in e.g., \cite{McGough:2016lol,Aharony:2018bad,Gorbenko:2018oov,Conti:2019dxg,Guica:2019nzm,Jorjadze:2020ili}
(for a review and further references see also \cite{Jiang:2019epa}).}. Remarkably, no limiting process is involved in the nonlinear map
between the conformal algebra and its ultra/non-relativistic version.
This map, and its corresponding deformation were then shown to be
recovered through a class of ``infinite boosts'' spanned by certain
degenerate (non-invertible) linear transformations acting on the coordinates
\cite{Bagchi:2022nvj}. Further recent interesting results about continuous
$\sqrt{T\bar{T}}$ deformations have also been addressed for field
theories in 2D along different approaches and points of view in \cite{Conti:2022egv},
\cite{Ferko:2022lol}, \cite{Babaei-Aghbolagh:2022kfz} (see also
\cite{Hou:2022csf}, \cite{Borsato:2022tmu}), and for the deformation
of two free harmonic oscillators in classical mechanics \cite{Garcia:2022wad}. 

One of the main purposes of our work is showing that the conformal
algebra in 2D admits a ``nonlinear $SO(1,1)$ automorphism'' that
maps the algebra to itself, which once implemented on generic classical
CFT$_{2}$'s, continuous $\sqrt{T\bar{T}}$ deformations naturally
emerge; as well as exploring some of its dynamical, geometric, thermodynamic
and limiting features. 

\section{Nonlinear $SO(1,1)$ automorphism \label{sec:NonlinearSO11Automorphism}}

The Poisson brackets of the generators of the conformal algebra in
2D, spanned by two copies of the Witt (centerless Virasoro) algebra,
fulfill
\begin{eqnarray}
\left\{ T\left(\phi\right),T\left(\varphi\right)\right\}  & = & \left(2T\left(\phi\right)\partial_{\phi}+\partial_{\phi}T\left(\phi\right)\right)\delta\left(\phi-\varphi\right)\;,\nonumber \\
\left\{ \bar{T}\left(\phi\right),\bar{T}\left(\varphi\right)\right\}  & = & -\left(2\bar{T}\left(\phi\right)\partial_{\phi}+\partial_{\phi}\bar{T}\left(\phi\right)\right)\delta\left(\phi-\varphi\right)\;,\label{eq:Conformal-Algebra-Cont}
\end{eqnarray}
with $\left\{ T\left(\phi\right),\bar{T}\left(\varphi\right)\right\} =0$.
Thus, Poisson brackets of functionals of the form $F=F[T,\bar{T}]$
can be directly computed in terms of the ``fundamental'' brackets
in \eqref{eq:Conformal-Algebra-Cont}. One can then verify that the
following nonlinear relations
\begin{align}
T_{\alpha} & =T\cosh^{2}\left(\frac{\alpha}{2}\right)+\bar{T}\sinh^{2}\left(\frac{\alpha}{2}\right)-\sqrt{T\bar{T}}\sinh\left(\alpha\right)\;\;,\nonumber \\
\bar{T}_{\alpha} & =\bar{T}\cosh^{2}\left(\frac{\alpha}{2}\right)+T\sinh^{2}\left(\frac{\alpha}{2}\right)-\sqrt{T\bar{T}}\sinh\left(\alpha\right)\;\;,\label{eq:NonlinearMap}
\end{align}
where $\alpha$ is a dimensionless constant, are such that $T_{\alpha}$
and $\bar{T}_{\alpha}$ precisely obey the same conformal algebra
of $T$ and $\bar{T}$ in \eqref{eq:Conformal-Algebra-Cont}. In order
to carry out an explicit checking, it is useful to note that $\sqrt{T}$
and $\sqrt{\bar{T}}$ behave as $U(1)$ currents, and in terms that
currents, the map in \eqref{eq:NonlinearMap} can be seen as a standard
(linear) $SO(1,1)$ automorphism of their algebra (see section \ref{sec:OverviewEndingRemarks}). 

Therefore, eq. \eqref{eq:NonlinearMap} provides the searched for
nonlinear map of the conformal algebra \eqref{eq:Conformal-Algebra-Cont}
to itself, that can be regarded as a nonlinear $SO(1,1)$ automorphism. 

It is also useful to change the basis of the conformal algebra \eqref{eq:Conformal-Algebra-Cont}
in terms of the energy and momentum densities, given by $H=T+\bar{T}$
and $J=T-\bar{T}$, respectively. In this basis, the nonlinear automorphism
\eqref{eq:NonlinearMap} manifestly preserves the form of the momentum
density, since
\begin{equation}
J=T-\bar{T}=T_{\alpha}-\bar{T}_{\alpha}\,,
\end{equation}
and naturally induces a flow on the energy density by a continuous
marginal $\sqrt{T\bar{T}}$ deformation, given by $H_{\alpha}=T_{\alpha}+\bar{T}_{\alpha}=\cosh\left(\alpha\right)\left(T+\bar{T}\right)-2\sinh\left(\alpha\right)\sqrt{T\bar{T}}$,
or equivalently
\begin{align}
H_{\alpha} & =\cosh\left(\alpha\right)H-\sinh\left(\alpha\right)\sqrt{H^{2}-J^{2}}\;.\label{eq:Palpha}
\end{align}
It also worth pointing out that this nonlinear automorphism is the
most general one that preserves the momentum density $J$. Indeed,
one can prove that for a generic nonlinear relationship of the form
$\tilde{H}=\tilde{H}\left[H,J\right]$, requiring the deformed generators
$J$ and $\tilde{H}$ to fulfill the conformal algebra fixes $\tilde{H}=H_{\alpha}$
as in \eqref{eq:Palpha}.

\section{Continuous $\sqrt{T\bar{T}}$ deformations}

As a direct consequence of the nonlinear automorphism \eqref{eq:Palpha},
if the energy density of a CFT$_{2}$ is replaced by the ``spectrally
flowed'' one in \eqref{eq:Palpha}, i.e., $H\rightarrow H_{\alpha}$,
the corresponding deformed action $I_{\alpha}$ necessarily retains
the conformal symmetry. The deformation of a generic CFT$_{2}$ action
can then be readily implemented in Hamiltonian form, which in the
conformal gauge explicitly reads 
\begin{equation}
I_{\alpha}=\int d^{2}x\,L_{\alpha}=\int d^{2}x\,\left(\Pi\dot{\Phi}-H_{\alpha}\right)\,,\label{eq:I_alpha}
\end{equation}
where $\Phi$ and $\Pi$ collectively denote the fields and their
momenta.

It is also worth noting that, since the marginally deformed energy
density $H_{\alpha}$ in \eqref{eq:Palpha} identically fulfills
\begin{equation}
\frac{\partial H_{\alpha}}{\partial\alpha}=\sqrt{H_{\alpha}^{2}-J^{2}}\;,\label{eq:Flow_NonlinearMap}
\end{equation}
the deformed Lagrangian satisfies the following flow equation
\begin{equation}
\frac{\partial L_{\alpha}}{\partial\alpha}=\sqrt{H_{\alpha}^{2}-J^{2}}=2\sqrt{T_{\alpha}\bar{T}_{\alpha}}=\sqrt{\det\left(T_{\mu\nu}^{\left(\alpha\right)}\right)}\;,\label{eq:Flow_Lagrangian}
\end{equation}
where $T_{\mu\nu}^{\left(\alpha\right)}$ is the stress-energy tensor
of the deformed CFT$_{2}$. Thus, eq. \eqref{eq:Flow_Lagrangian}
precisely agrees with the flow that has been recently considered in
\cite{Conti:2022egv}, \cite{Ferko:2022lol} and \cite{Babaei-Aghbolagh:2022kfz}
(see also \cite{Hou:2022csf}, \cite{Borsato:2022tmu}), for some
concrete examples in the Lagrangian formalism. Indeed, in \cite{Conti:2022egv}
it was shown that a dimensional reduction of the ModMax theory \cite{Bandos:2020jsw}
(see also \cite{Kosyakov:2020wxv,Sorokin:2021tge}) to two spacetime
dimensions corresponds to the continuous $\sqrt{T\bar{T}}$ deformation
of $N=2$ free bosons, and then extended the deformation to arbitrary
$N$. The deformation of free bosons was also addressed in \cite{Ferko:2022lol}
from a more general approach, where it was argued that the deformation
should also work well for a wider class of examples, as confirmed
in \cite{Borsato:2022tmu}. The same result for free bosons was also
obtained in \cite{Babaei-Aghbolagh:2022kfz} from a perturbative approach.
Later on, in \cite{Hou:2022csf} it was shown that the flow equation
\eqref{eq:Flow_NonlinearMap} in Lagrangian form can be reduced to
a set of first order non-linear PDE's by method of characteristics,
and applied them to recover the deformation of free bosons.

We should highlight that one of the advantages of the Hamiltonian
approach is that the general solution of the Lagrangian flow equation
\eqref{eq:Flow_Lagrangian} can be analytically solved in closed form,
recovering the nonlinear automorphism of the conformal algebra. Indeed,
solving \eqref{eq:Flow_Lagrangian} in Hamiltonian form amounts to
find the general solution of \eqref{eq:Flow_NonlinearMap}, which
is given by

\begin{equation}
H_{\alpha}=\left|J\right|\cosh\left(\alpha-f\left(H,J\right)\right)\;,\label{eq:fflow}
\end{equation}
with $f$ an arbitrary function of $H$ and $J$, whose precise form
becomes fixed as 
\begin{equation}
f\left(H,J\right)=\cosh^{-1}\left(\frac{H}{\left|J\right|}\right)\;,\label{eq:f_initCond}
\end{equation}
once the initial condition $H_{0}=H$ is imposed. Therefore, the energy
flow $H_{\alpha}$ in \eqref{eq:Palpha} is recovered once \eqref{eq:f_initCond}
is replaced into \eqref{eq:fflow}.

As a simple explicit example, the continuous $\sqrt{T\bar{T}}$ deformation
of $N$ free bosons with flat target metric
\begin{equation}
I_{0}\left[\Phi^{I}\right]=-\frac{1}{2}\int d^{2}x\sqrt{-g}\delta_{IK}\partial_{\mu}\Phi^{I}\partial^{\mu}\Phi^{K}\ ,\label{eq:I-NfreeBs-CFT-1}
\end{equation}
can then be readily implemented just by replacing the energy density
$H=\frac{1}{2}\left(\Pi^{I}\Pi_{I}+\Phi^{\prime I}\Phi_{I}^{\prime}\right)$
by $H_{\alpha}$ in \eqref{eq:Palpha}, where $\Pi_{I}=\frac{\delta L}{\delta\dot{\Phi}^{I}}$
and $J=\Pi_{I}\Phi^{I\prime}$, so that the deformed Hamiltonian action
in the conformal gauge \eqref{eq:I_alpha} in this case reads
\begin{equation}
I_{\alpha}\left[\Phi^{I},\Pi_{J}\right]=\int dx^{2}\left(\Pi_{I}\dot{\Phi}^{I}-H_{\alpha}\right)\ ,\label{eq:Ialpha-N}
\end{equation}
with
\begin{equation}
H_{\alpha}=\frac{1}{2}\left(\Pi^{I}\Pi_{I}+\Phi^{\prime I}\Phi_{I}^{\prime}\right)\cosh\left(\alpha\right)-\sinh\left(\alpha\right)\sqrt{\frac{1}{4}\left(\Pi^{I}\Pi_{I}+\Phi^{\prime I}\Phi_{I}^{\prime}\right)^{2}-\left(\Pi_{I}\Phi^{I\prime}\right)^{2}}\;.\label{eq:Palpha-N}
\end{equation}
Note that, as it also occurs for the ModMax theory \cite{Bandos:2020jsw}
(see also \cite{Escobar:2021mpx}), the equivalence of the deformed
Lagrangian action for the free bosons obtained in \cite{Conti:2022egv},
\cite{Ferko:2022lol}, \cite{Babaei-Aghbolagh:2022kfz} and also in
\cite{Hou:2022csf} with the Hamiltonian action given by \eqref{eq:Ialpha-N}
with \eqref{eq:Palpha-N}, requires careful implementation of a suitable
Legendre transformation. In the case of a single free boson ($N=1$)
the deformation is trivial because the spectrally flowed energy simplifies
as $H_{\alpha}=\frac{1}{2}\left(\Pi^{2}e^{-\alpha}+e^{\alpha}\Phi^{\prime2}\right)$,
so that the momentum is obtained from its own field equation $\Pi=e^{\alpha}\dot{\Phi}$.
Thus, substituting in \eqref{eq:Ialpha-N}, the deformed action is
just a rescaling of \eqref{eq:I-NfreeBs-CFT-1}, given by $I_{\alpha}=e^{\alpha}I_{0}$.

A manifestly covariant form of the continuously $\sqrt{T\bar{T}}$-deformed
action is briefly discussed in section \ref{sec:OverviewEndingRemarks}.

\section{Geometric realization\label{sec:GeometricRealization}}

Following the lines of \cite{Rodriguez:2021tcz} one can show that
the continuous $\sqrt{T\bar{T}}$ deformation can be geometrically
implemented in terms of the original (undeformed) CFT$_{2}$ on a
precise field-dependent Riemannian metric. It is then useful and instructive
to consider the undeformed theory in a generic (non-conformal) gauge,
so that in a local patch, the two-dimensional background metric belongs
to the conformal class of the following one 
\begin{equation}
ds^{2}=-N^{2}dt^{2}+\left(d\phi+N^{\phi}dt\right)^{2}\ ,\label{eq:ds2-N-Nphi-1}
\end{equation}
where $N$, $N^{\phi}$ are the lapse and shift functions, respectively.
Since the total Hamiltonian of the undeformed CFT$_{2}$ is now given
by
\begin{align}
\mathscr{H}_{0} & =\int d\phi\left(\mu T+\bar{\mu}\bar{T}\right)=\int d\phi\left(NH+N^{\phi}J\right)\ ,\label{eq:H0}
\end{align}
with $\mu=N+N^{\phi}$ and $\bar{\mu}=N-N^{\phi}$, that of deformed
theory is obtained through the replacement $H\rightarrow H_{\alpha}$,
which reads 
\begin{equation}
\mathscr{H}_{\alpha}=\int d\phi\left(\mu T_{\alpha}+\bar{\mu}\bar{T}_{\alpha}\right)=\int d\phi\left(NH_{\alpha}+N^{\phi}J\right)\ .\label{eq:Halpha}
\end{equation}
The conservation laws then acquire the following form
\begin{align}
\dot{T}_{\alpha} & =\left\{ T_{\alpha},\mathscr{H}_{\alpha}\right\} =2T_{\alpha}\mu^{\prime}+\mu T_{\alpha}^{\prime}\,,\\
\dot{\bar{T}}_{\alpha} & =\left\{ \bar{T}_{\alpha},\mathscr{H}_{\alpha}\right\} =-\left(2\bar{T}_{\alpha}\bar{\mu}^{\prime}+\bar{\mu}\bar{T}_{\alpha}^{\prime}\right)\,,
\end{align}
and in absence of global obstructions, the canonical generators can
be written as
\begin{equation}
{\cal Q}_{\alpha}\left[\eta,\bar{\eta}\right]=\int d\phi\left(\eta T_{\alpha}+\bar{\eta}\bar{T}_{\alpha}\right)\,,\label{eq:Qalpha}
\end{equation}
being conserved ($\dot{{\cal Q}}_{\alpha}=0$) provided the parameters
fulfill
\begin{align}
\dot{\eta} & =\mu\eta^{\prime}-\eta\mu^{\prime}\;\;,\;\;\dot{\bar{\eta}}=\bar{\eta}\bar{\mu}^{\prime}-\bar{\mu}\bar{\eta}^{\prime}\,.\label{eq:eta_punto}
\end{align}
The transformation laws of $T_{\alpha}$ and $\bar{T}_{\alpha}$ then
follow from $\delta T_{\alpha}=\left\{ T_{\alpha},{\cal Q}_{\alpha}\left[\eta\right]\right\} $
and $\delta\bar{T}_{\alpha}=\left\{ \bar{T}_{\alpha},{\cal Q}_{\alpha}\left[\bar{\eta}\right]\right\} $,
which in terms of null coordinates, $x=t+\phi$ and $\bar{x}=t-\phi$,
read 
\begin{equation}
\delta T_{\alpha}=2T_{\alpha}\partial\eta+\partial T_{\alpha}\eta\;\;,\;\;\delta\bar{T}_{\alpha}=2\bar{T}_{\alpha}\bar{\partial}\bar{\eta}+\bar{\partial}\bar{T}_{\alpha}\bar{\eta}\ .\label{eq:deltaT_alpha}
\end{equation}
Note that in the conformal gauge ($N=1$, $N^{\phi}=0$) the independent
components of the stress energy tensor and the parameters become (anti-)chiral,
i.e., $\dot{T}_{\alpha}=T_{\alpha}^{\prime}$, $\dot{\bar{T}}_{\alpha}=-\bar{T}_{\alpha}^{\prime}$;
$\dot{\eta}=\eta^{\prime}$, $\dot{\bar{\eta}}=-\bar{\eta}^{\prime}$. 

A geometric description of the continuous $\sqrt{T\bar{T}}$ deformation
then follows from the fact that the deformed total Hamiltonian \eqref{eq:Halpha}
is a homogeneous functional of degree one in $T$ and $\bar{T}$ (as
well as in $J$, $H$) so that it fulfills 
\begin{equation}
\mathscr{H}_{\alpha}=\int d\phi\left(\frac{\delta\mathscr{H}_{\alpha}}{\delta T}T+\frac{\delta\mathscr{H}_{\alpha}}{\delta\bar{T}}\bar{T}\right)=\int d\phi\left(\frac{\delta\mathscr{H}_{\alpha}}{\delta H}H+\frac{\delta\mathscr{H}_{\alpha}}{\delta J}J\right)\ .\label{eq:HalphaTTPJ}
\end{equation}
Therefore, comparing $\mathscr{H}_{\alpha}$ in \eqref{eq:HalphaTTPJ}
and \eqref{eq:H0}, it is apparent that the deformed theory becomes
equivalently described by the original CFT$_{2}$ on a field-dependent
curved metric, whose lapse and shift functions are respectively given
by $N_{\alpha}=\frac{\delta\mathscr{H}_{\alpha}}{\delta H}$ and $N_{\alpha}^{\phi}=\frac{\delta\mathscr{H}_{\alpha}}{\delta J}$;
i.e., 
\begin{equation}
ds_{\left(\alpha\right)}^{2}=-\left(\frac{\delta\mathscr{H}_{\alpha}}{\delta H}\right)^{2}dt^{2}+\left[d\phi+\left(\frac{\delta\mathscr{H}_{\alpha}}{\delta J}\right)dt\right]^{2}\;,
\end{equation}
which in terms of $T_{\alpha}$ and $\bar{T}_{\alpha}$ reads
\begin{align}
ds_{\left(\alpha\right)}^{2} & =-\left[\cosh\left(\alpha\right)+\frac{T_{\alpha}+\bar{T}_{\alpha}}{2\sqrt{T_{\alpha}\bar{T}_{\alpha}}}\sinh\left(\alpha\right)\right]^{-2}N^{2}dt^{2}\nonumber \\
 & +\left[d\phi+\left(N^{\phi}+\frac{\left(T_{\alpha}-\bar{T}_{\alpha}\right)\tanh\left(\alpha\right)}{\left(T_{\alpha}+\bar{T}_{\alpha}\right)\tanh\left(\alpha\right)+2\sqrt{T_{\alpha}\bar{T}_{\alpha}}}N\right)dt\right]^{2}\;,\label{eq:dS2-TT}
\end{align}
where $N$ and $N^{\phi}$ stand for the (field-independent) lapse
and shift functions of the original undeformed background metric in
\eqref{eq:ds2-N-Nphi-1}.\footnote{As in \cite{Rodriguez:2021tcz}, the Ricci scalar of the field-dependent
metric $g_{\mu\nu}^{\left(\alpha\right)}$ is not diffeomorphic to
that of the undeformed one $g_{\mu\nu}$ ($R^{\left(\alpha\right)}\neq R$).
Following another approach, and in the context of $T\bar{T}$-like
deformations, different field-dependent modifications of the background
metric have also been proposed in \cite{Conti:2022egv}. In contradistinction,
note that in the geometric interpretation of the $T\bar{T}$ deformations,
the corresponding background metrics turn out to be related through
field-dependent diffeomorphisms \cite{Dubovsky:2017cnj,Cardy:2018sdv,Dubovsky:2018bmo,Conti:2018tca}.}.

Note that once the theory is deformed, the manifest field dependence
of the lapse and shift functions $N_{\alpha}$ and $N_{\alpha}^{\phi}$
of the deformed metric \eqref{eq:dS2-TT} yields a local obstruction
that prevents to gauge them away.

It is also worth highlighting that the field-dependent Riemannian
metric $g_{\mu\nu}^{\left(\alpha\right)}$ and the conserved stress-energy
tensor $T_{\;\mu\nu}^{\left(\alpha\right)}$, become inextricably
intertwined. Thus, since both structures are field dependent, their
functional variation
\begin{equation}
\delta_{\xi}g_{\mu\nu}^{\left(\alpha\right)}=\frac{\delta g_{\mu\nu}^{\left(\alpha\right)}}{\delta T_{\alpha}}\delta_{\xi}T_{\alpha}+\frac{\delta g_{\mu\nu}^{\left(\alpha\right)}}{\delta\bar{T}_{\alpha}}\delta_{\xi}\bar{T}_{\alpha}\;\;,\;\;\delta_{\xi}T_{\;\mu\nu}^{\left(\alpha\right)}=\frac{\delta T_{\;\mu\nu}^{\left(\alpha\right)}}{\delta T_{\alpha}}\delta_{\xi}T_{\alpha}+\frac{\delta T_{\;\mu\nu}^{\left(\alpha\right)}}{\delta\bar{T}_{\alpha}}\delta_{\xi}\bar{T}_{\alpha}\;,\label{eq:FunctionalVariations-g-theta-tilde-1-1}
\end{equation}
has to be taken into account when acting on them with diffeomorphisms
$\xi=\xi^{\mu}\partial_{\mu}$. Hence, the symmetries of the deformed
theory can be seen to geometrically arise from diffeomorphisms preserving
both relevant structures up to local scalings, which yields to \textit{deformed
conformal Killing equations}, given by

\[
\nabla_{\mu}^{\left(\alpha\right)}\xi_{\nu}+\nabla_{\nu}^{\left(\alpha\right)}\xi_{\mu}-\lambda g_{\mu\nu}^{\left(\alpha\right)}=\delta_{\xi}g_{\mu\nu}^{\left(\alpha\right)}\ ,
\]
\begin{equation}
\mathcal{L}_{\xi}T_{\;\mu\nu}^{\left(\alpha\right)}=\delta_{\xi}T_{\;\mu\nu}^{\left(\alpha\right)}\ ,\label{Deformed-Conf. -Killing equations-1}
\end{equation}
where $\nabla_{\mu}^{\left(\alpha\right)}$ is the covariant derivative
with respect to the metric $g_{\mu\nu}^{\left(\alpha\right)}$, and
$\mathcal{L}_{\xi}$ stands for the Lie derivative.

Noteworthy, the deformed conformal Killing equations \eqref{Deformed-Conf. -Killing equations-1}
can be exactly solved from scratch for the deformed metric and stress-energy
tensor, $g_{\mu\nu}^{\left(\alpha\right)}$ and $T_{\;\mu\nu}^{\left(\alpha\right)}$,
whose solution is given by diffeomorphisms of the form
\begin{equation}
\xi^{\mu}=N^{-1}\left(\eta_{H},N\eta_{J}-N^{\phi}\eta_{H}\right)\,,
\end{equation}
with parameters $\eta=\eta_{H}+\eta_{J}$ and $\bar{\eta}=\eta_{H}-\eta_{J}$,
that fulfill the same eq. \eqref{eq:eta_punto} above, while the transformation
laws of $T_{\alpha}$ and $\bar{T}_{\alpha}$ are found to be given
by the expected ones as in \eqref{eq:deltaT_alpha}. It is then amusing
to verify that the standard conformal Killing vectors emerge from
solving \eqref{Deformed-Conf. -Killing equations-1} on the field-dependent
deformed metric $g_{\mu\nu}^{\left(\alpha\right)}$ in \eqref{eq:dS2-TT},
due to the fact that the conformal Killing equation has been suitably
deformed. Therefore, the transformation law of the fields in the deformed
theory agrees with those of the original undeformed (primary) fields.
Indeed, collectively denoting the fields by $\Phi$$,$ and expressing
them in covariant form, the conformal symmetries of the deformed theory
become just spanned by $\delta_{\xi}\Phi=\mathcal{L}_{\xi}\Phi$.

\section{Overview and ending remarks \label{sec:OverviewEndingRemarks}}

Since the nonlinear automorphism \eqref{eq:NonlinearMap} preserves
the conformal symmetries, its corresponding continuous $\sqrt{T\bar{T}}$
deformation also yields to a CFT$_{2}$, and hence the deformation
keeps the integrability properties of the undeformed theory. Furthermore,
as pointed out in section \ref{sec:NonlinearSO11Automorphism}, $\sqrt{T}$
and $\sqrt{\bar{T}}$ behave as currents, because their corresponding
Poisson brackets fulfill
\begin{align}
\left\{ \sqrt{T\left(\phi\right)},T\left(\varphi\right)\right\}  & =\partial_{\phi}\left(\sqrt{T\left(\phi\right)}\delta\left(\phi-\varphi\right)\right)\;,\\
\left\{ \sqrt{T\left(\phi\right)},\sqrt{T\left(\varphi\right)}\right\}  & =\frac{1}{2}\partial_{\varphi}\delta\left(\phi-\varphi\right)\;,
\end{align}
and similarly for $\sqrt{\bar{T}}$ with a corresponding minus sign
on the r.h.s. Thus, $\sqrt{T}$ and $\sqrt{\bar{T}}$ can be naturally
assembled as the components of a vector
\begin{equation}
j^{I}=\left(\begin{array}{c}
\sqrt{T}\\
\sqrt{\bar{T}}
\end{array}\right)\;,\label{eq:J_I}
\end{equation}
so that
\begin{equation}
\left\{ j^{I}\left(\phi\right),j^{J}\left(\varphi\right)\right\} =-\frac{1}{2}\eta^{IJ}\partial_{\varphi}\delta\left(\phi-\varphi\right)\;,
\end{equation}
with $\eta^{IJ}=diag(-1,1)$. Note that in terms of the currents $j^{I}$,
the map \eqref{eq:NonlinearMap} is equivalently expressed as
\begin{equation}
j_{\left(\alpha\right)}^{I}=\Lambda_{\,K}^{I}\left(\frac{\alpha}{2}\right)j^{K}\;,
\end{equation}
with 
\begin{equation}
\Lambda_{\,K}^{I}\left(\alpha\right)=\left(\begin{array}{cc}
\cosh\left(\alpha\right) & -\sinh\left(\alpha\right)\\
-\sinh\left(\alpha\right) & \cosh\left(\alpha\right)
\end{array}\right)\,,
\end{equation}
and therefore, the nonlinear automorphism in \eqref{eq:NonlinearMap}
can be seen to emerge from a standard (linear) $SO(1,1)$ one, through
a sort of ``inverse Sugawara construction''. 

The currents $j^{I}$ are also useful in order to construct manifestly
invariant objects under the nonlinear automorphism \eqref{eq:NonlinearMap},
as it is the case of the momentum density, since it can be expressed
as 
\begin{equation}
J=T-\bar{T}=-\eta_{IJ}j^{I}j^{J}\,.
\end{equation}
In this sense, it is also worth pointing out that for negative values
of $T$ and $\bar{T}$, consistency of the map requires using the
negative branch of the square root in \eqref{eq:NonlinearMap}. Hence
for $T=-\mathcal{T}$ and $T=-\overline{\mathcal{T}}$, with $\mathcal{T},\overline{\mathcal{T}}>0,$
the corresponding currents assemble as components of a contravariant
vector, i.e., 
\begin{equation}
\mathcal{J}_{I}=\left(\begin{array}{cc}
\sqrt{-\mathcal{T}} & \sqrt{-\overline{\mathcal{T}}}\end{array}\right),
\end{equation}
because they transform according to 
\begin{equation}
\mathcal{J}_{I}^{\alpha}=\mathcal{J}_{K}\Lambda_{\;I}^{K}\left(\dfrac{\alpha}{2}\right)\;.
\end{equation}
 Besides, the centrally extended conformal algebra, described by two
copies of the Virasoro algebra, can also be seen to admit a nontrivial
automorphism that is necessarily nonlocal. Nonetheless, the nonlinear
automorphism \eqref{eq:NonlinearMap} still holds when only the zero
modes are involved in the map. Consequently, Cardy formula can be
seen to be invariant under the nonlinear automorphism. Indeed, once
expressed in terms of the (negative) left and right ground state energies,
denoted by $\mathcal{T}\text{and }\overline{\mathcal{T}}$, respectively,
the asymptotic growth of the number of states can be written as
\begin{equation}
S=4\pi\left(\sqrt{-\mathcal{T}T}+\sqrt{-\overline{\mathcal{T}}\bar{T}}\right)=4\pi\;\mathcal{J}_{I}j^{I}=4\pi\;\mathcal{J}_{I}^{\alpha}j_{\alpha}^{I}=4\pi\left(\sqrt{-\mathcal{T}_{\alpha}T_{\alpha}}+\sqrt{-\overline{\mathcal{T}}_{\alpha}\bar{T}_{\alpha}}\right)\;,
\end{equation}
being manifestly invariant under the nonlinear $SO(1,1)$ automorphism.

Furthermore, the nonlinear map between the conformal algebra in 2D
and its ultra/non-relativistic version recently found in \cite{Rodriguez:2021tcz},
including the corresponding finite $\sqrt{T\bar{T}}$ deformation
from a CFT$_{2}$ to a field theory invariant under BMS$_{3}$, can
be recovered from the nonlinear automorphism in a limiting case. This
can be seen as follows: rescaling $H_{\alpha}$ in \eqref{eq:Palpha}
as
\begin{align}
P_{\alpha}=\frac{H_{\alpha}}{\cosh\left(\alpha\right)} & =H-\tanh\left(\alpha\right)\sqrt{H^{2}-J^{2}}\;,\label{eq:Ptilde_alpha}
\end{align}
one can perform an In{\"o}n{\"u}-Wigner contraction of conformal algebra spanned
by $J$ and $P_{\alpha}$, so that in the limit $\alpha\rightarrow\pm\infty$,
both branches of the supertranslation generators of the map 
\begin{equation}
P_{\left(\pm\right)}=H\mp\sqrt{H^{2}-J^{2}}\;,\label{eq:Ptildemas-menos}
\end{equation}
clearly fulfill BMS$_{3}$ algebra; in full agreement with the result
in \cite{Rodriguez:2021tcz} that was directly obtained from \eqref{eq:Ptildemas-menos}
without any sort of limiting process. Note that since the energy density
rescales as in \eqref{eq:Ptilde_alpha}, consistency with time evolution
implies that the time coordinate rescales according to $\tilde{t}=t\cosh\left(\alpha\right)$,
so that in terms of $\text{\ensuremath{\tilde{t}}}$ the deformed
action reads
\begin{equation}
I_{\alpha}=I_{0}-2\tanh\left(\alpha\right)\int d\tilde{t}d\phi\sqrt{T\bar{T}}\,.
\end{equation}
Thus, before taking the limit, and for a generic gauge choice, the
continuously $\sqrt{T\bar{T}}$-deformed deformed action can be written
in a manifestly covariant way as 

\begin{equation}
I_{\alpha}=I_{0}-\tanh\left(\alpha\right)\int d^{2}\tilde{x}\sqrt{\det T_{\mu\nu}}\;,
\end{equation}
where $T_{\mu\nu}$ stands for stress-energy tensor of the undeformed
CFT$_{2}$, and it is implicitly assumed that $I_{0}$ is given in
Hamiltonian form.

Hence, in the limit $\alpha\rightarrow\pm\infty$$,$ the finite $\sqrt{T\bar{T}}$
deformation in \cite{Rodriguez:2021tcz}, given by 
\begin{equation}
\tilde{I}=I_{0}\pm\int d^{2}\tilde{x}\sqrt{\det T_{\mu\nu}}\;,
\end{equation}
is recovered. From this procedure once concludes that that the finite
$\sqrt{T\bar{T}}$ deformation necessarily yields to an ultra-relativistic
theory, since the speed of light can be identified as $c=1/\cosh\left(\alpha\right)$. 

An interesting remark that concerns the pure BMS$_{3}$ algebra is
in order. Note that each branch of the nonlinear map from the conformal
algebra in 2D to BMS$_{3}$ in \eqref{eq:Ptildemas-menos} reproduces
precisely the same algebra. Thus, since both branches are related
as

\begin{equation}
P_{\left(+\right)}=\frac{J^{2}}{P_{\left(-\right)}}\,,\label{eq:DiscreteAutomorpBMS3}
\end{equation}
this last equation provides a nontrivial discrete nonlinear automorphism
of the BMS$_{3}$ algebra. Indeed, one can prove that this is the
most general nonlinear automorphism of the form $\tilde{P}=\tilde{P}\left[P,J\right]$
that preserves superrotations $J$ (up to a trivial constant rescaling
of supertranslations). The discrete automorphism \eqref{eq:DiscreteAutomorpBMS3}
appears to be deeply related to the inequivalent kinds of Carrollian
limits discussed in \cite{Henneaux:2021yzg}, dubbed as those of electric
and magnetic type (see also \cite{Duval:2014uoa}). Indeed, for a
single scalar field, the corresponding deformation of the action associated
to the nonlinear discrete automorphism of BMS$_{3}$ in \eqref{eq:DiscreteAutomorpBMS3}
maps the Carrollian electric-like action to the magnetic-like one
and vice-versa. In this sense, \eqref{eq:DiscreteAutomorpBMS3} could
be regarded as a duality relation between Carrollian theories of electric
and magnetic type. 

As an ending remark, we point out that the nonlinear $SO(1,1)$ automorphism
of the conformal algebra in 2D can be extended to a three-parameter
$ISO(1,1)$ one, given by 
\begin{align}
T_{\alpha,\eta} & =\left[\sqrt{T}\cosh\left(\frac{\alpha}{2}\right)-\sqrt{\bar{T}}\sinh\left(\frac{\alpha}{2}\right)+\eta\right]^{2}\;,\nonumber \\
\bar{T}_{\alpha,\bar{\eta}} & =\left[\sqrt{\bar{T}}\cosh\left(\frac{\alpha}{2}\right)-\sqrt{T}\sinh\left(\frac{\alpha}{2}\right)+\bar{\eta}\right]^{2}\;,
\end{align}
with $\alpha$, $\eta$ and $\bar{\eta}$ constants. Note that in
terms of the currents \eqref{eq:J_I} the extended automorphism reads
\begin{equation}
j_{\left(\alpha,\eta,\bar{\eta}\right)}^{I}=\Lambda_{\;J}^{I}\left(\frac{\alpha}{2}\right)j^{J}+\eta^{I}\;,
\end{equation}
with 
\begin{equation}
\eta^{I}=\left(\begin{array}{c}
\eta\\
\bar{\eta}
\end{array}\right)\,,
\end{equation}
which clearly corresponds to a standard linear automorphism. Three-parameter
deformations of CFT$_{2}$'s can then be readily performed by virtue
of the extended nonlinear automorphism. Nonetheless, the momentum
density is no longer preserved, and the energy density ceases to be
homogeneous of degree one, which precludes a direct geometric interpretation
of the 3-parameter deformation of the action, as that performed in
section \ref{sec:GeometricRealization}. Further details about the
extended automorphism, as well as its application in the context of
celestial holography (see e.g., \cite{Pasterski:2016qvg,Pasterski:2017kqt,Pasterski:2017ylz}),
are expected to be discussed elsewhere.

\section*{Acknowledgments}

We thank Hern\'{a}n Gonz\'{a}lez, Marc Henneaux, Pulastya Parekh, Alfredo
P\'{e}rez, Miguel Pino and Pablo Rodr\'{i}guez for useful comments and discussions.
This research has been partially supported by ANID FONDECYT grants
N$^{\circ}$ 1211226, 1220910 and 1221624.

\end{document}